\pdfoutput=1
\documentclass[a4paper,11pt]{article}
\usepackage{amsmath}
\usepackage{graphicx}
\usepackage{amssymb}
\usepackage{amsfonts}
\usepackage{latexsym}
\usepackage{color}

\newcommand{\rL}{\rho_{\Lambda\text{eff}}}
\newcommand{\pL}{p_{\Lambda\text{eff}}}
\newcommand{\GB}{{\mathcal G}}
\newcommand{\CC}{\rho_{\Lambda}}

\begin{document}

\begin{center}

\phantom{} \vspace{1.5cm}
{\Large The \textit{Relaxed Universe}: towards solving the cosmological
constant problem dynamically from an effective action functional of
gravity} \vskip 1cm

{\bf Florian Bauer\footnote{fbauer@ecm.ub.es}, Joan Sol\`a}\footnote{sola@ecm.ub.es}\\

HEP Group, Dept.\ ECM and Institut de Ci{\`e}ncies del Cosmos\\
Univ.\ de Barcelona, Av.\ Diagonal 647, E-08028 Barcelona, Catalonia,
Spain

\vspace{0.5cm}

\quad { \bf Hrvoje
 \v{S}tefan\v{c}i\'{c}}\footnote{shrvoje@thphys.irb.hr} \\

Theoretical Physics Division, Rudjer Bo\v{s}kovi\'{c}
Institute \\PO Box 180, HR-10002 Zagreb, Croatia\\

\vspace{0.5cm}

\vskip 2cm
\end{center}

\begin{abstract}
We present an unconventional approach for addressing the old cosmological constant (CC) problem in a class of $F(R,\GB)$ models of modified gravity. For a CC of
arbitrary size and sign the corresponding cosmological evolution
follows an expansion history which strikingly resembles that of our
real universe. The effects of the large CC are relaxed dynamically
and there is no fine-tuning at any stage. In addition, the
relaxation mechanism alleviates the coincidence problem. The upshot
is that a large cosmological constant and the observed cosmic
expansion history coexist peacefully in {\em the Relaxed Universe}.
This model universe can be thought of as an interesting
preliminary solution of the cosmological constant problem, in the
sense that it provides a successful dynamical mechanism able to
completely avoid the fine-tuning problem (the toughest aspect of the CC problem). However, since the Relaxed Universe is formulated
within the context of modified gravity, it may still suffer of some
of the known issues associated with these theories, and therefore it
can be viewed only as a toy-model proposal towards a final solution
of the CC problem.
\end{abstract}

\newpage

\section{\large\bf Introduction}

The triumph of modern physics is largely reflected in our ability to
understand the properties of the universe and its dynamics from the
fundamental physical theories. Yet there is a prediction of quantum
field theory and string theory which is grossly inconsistent with
the existence of the observed universe. The vacuum energy density
calculated in these theories can be so large in absolute value
($|\rho_{\Lambda}|\simeq M_X^4$, with $M_X$ around the Planck mass)
that the universe would either recollapse immediately after the Big
Bang or the formation of galaxies that we observe would be
prevented. This huge gap in our understanding of nature, known as
{\em the old cosmological constant problem}, is one of the most
important unresolved issues in fundamental physics
\cite{weinberg89,CCproblem}. The cosmological constant (CC) receives
numerous contributions from zero point energies of quantum fields
and phase transitions in the early universe, along with an arbitrary
additive constant as allowed by general covariance. These effects
are of very different orders of magnitude, and all of them are many
orders of magnitude larger than the value consistent with the
observations, $\rho_{\Lambda}^0 \approx (2.3 \times 10^{-3} \,
\text{eV})^4$. There is the possibility that these disparate
contributions might cancel to produce the observed value
$\rho_{\Lambda}^0$, but this would entail an unacceptably large
fine-tuning. Definitely, we have to find another, more natural,
solution.

There have been numerous attempts to solve the old CC problem, e.g.\
a recent one is~\cite{Demir}. Many, if not all of them, and
especially those based on dynamical scalar fields, still suffer from
severe fine-tuning troubles in one form or another~\cite{weinberg89}. Rather than replacing the CC by scalar fields, one may think of the CC term as
an effective quantity $\rL$ evolving with the expansion rate of the
universe, $H$. The approach to a dynamical CC in quantum field theory
has been recently re-emphasized in \cite{ShapSol09}, and it has
proven fruitful in tackling the cosmic coincidence
problem\,\cite{LXCDM}. Maybe it can also be useful to deal with the
old CC problem. Thus, instead of looking for the solution in the
matter sector, the possibility that it could have a gravitational
origin should be seriously investigated. This might provide a well
behaved ``effective CC term'' $\rL(H)$, viz.\ one which is very large
in the early universe ($H\simeq M_X$) but very small at present
($H\simeq H_0$). In this letter, we present action functionals of
modified gravity which incorporate a very large
$|\CC|\,(\gg\rho_{\Lambda}^0)$ and still result in a cosmological
expansion with a measurable $\rL(H)$ which is (very) close to the
dynamics of the benchmark $\Lambda$CDM cosmological model.
The striking feature of these modified gravity models is that they relax a large CC without fine-tuning. The gravitational mechanism
for equilibrating the effects of a large CC we refer to as {\em the
relaxation mechanism} whereas the cosmological models in which this
mechanism is implemented we call {\em the Relaxed Universe}.

\section{\large\bf Dynamical relaxation of the cosmological term}

An essential ingredient of the new universe is the feedback between
the dynamics of space-time and the functioning of the relaxation
mechanism. The very expansion owing to the existence of a large CC
sets the relaxation mechanism in motion. The effects of modified
gravity equilibrate (dynamically) the effects of the large CC. As a
result, a realistic universe with a standard sequence of radiation
dominated, matter dominated and accelerating epochs unfolds
automatically. The relaxation mechanism based on the component with
an inhomogeneous equation of state (EOS) was introduced in
\cite{Stefancic08}, while in \cite{Odin} it was shown that the
inhomogeneous EOS is an effective description of modified gravity
effects. The following step was to construct a viable cosmological
model.

In Ref.~\cite{BSS09a} we have proposed a scenario which achieves
this goal in a dynamical way, thereby avoiding fine-tuning. In a
flat FLRW cosmos, we introduced a dynamical vacuum component, which
counter-acts the effect of a large $|\CC|$. The total vacuum energy
density was found to be $\rL=\CC+\beta/B$, where $\beta$ is a
constant parameter, and the function $B$ representing the heart of
the relaxation mechanism will follow shortly. First, suppose that
$\CC\sim-M_{\text{EW}}^{4}$ (with $M_{\text{EW}}={\cal O}(100)$ GeV)
as given by the large (and negative) vacuum contribution from the
Higgs potential in the electro-weak phase transition. Without
counter-measures, the negative $\CC$ would let the universe quickly
collapse into a catastrophic Big Crunch. To prevent this, one could
artificially add a fine-tuned CC counter-term. Our relaxation
mechanism, though, will do the job automatically. Let us illustrate
it first with a toy model in the radiation era, where the
deceleration parameter~$q=-{\ddot{a}a}/{\dot{a}^{2}}$ is close to
but smaller than unity, $q\lesssim1$. In this epoch, consider
$\beta>0$ and $B=H^2(1-q)$ given in terms of~$q$ and the Hubble
rate~$H=\dot{a}/a$. Clearly, the mechanism will stabilize the
expansion, because the big $\CC<0$ wants to rapidly initiate the
collapse, thereby pushing the deceleration $q$ closer to~$1$.
However, $q$ can never reach~$1$ because the
second term $\beta/[H^2(1-q)]>0$ in $\rL$ will become sufficiently
large to compensate the negative CC, finally resulting in
$|\rL|\ll|\CC|$ --- the CC has been relaxed. And vice versa: if
$\beta/B>0$ would dominate over $\CC$, the value of $q$ would move
away from $q\approx1$ towards smaller values due to the large
positive vacuum energy. Consequently, $\beta/[H^2(1-q)]$ will
decrease and $q$ will reach a stable point where $\CC$ and $\beta/B$
compensate each other. The resulting universe will expand like a
radiation cosmos since small changes of $q$ around $q\simeq 1$ are
sufficient to keep both terms of $\rL$ in equilibrium.

\section{\large\bf A more realistic model}

To apply the CC relaxation mechanism to the whole Big Bang expansion
history, consider the more sophisticated B-function in $\rL$ above
\begin{equation}
B=\frac{2}{3}R^{2}+\frac{1}{2}\GB+(y\,
R)^{3},\label{eq:B(R,G)}\end{equation} involving only the Ricci
scalar $R=6H^{2}(1-q)$ and the Gau\ss-Bonnet term $\GB=-24H^{4}q$,
both computed in the FLRW metric. Thus, $B$ reads explicitly
\begin{equation}
B=24H^{4}\left(q-\frac{1}{2}\right)(q-2)+H^{6}\left[6\,y(1-q)\right]^{3}\,.\label{eq:B(H,q)}
\end{equation}
It is easy to see that the second term, proportional to
$H^{6}(1-q)^{3}$, will be responsible for the CC relaxation in the
radiation regime ($q\simeq 1$), where the Hubble rate was large.
Since, however, $H$ decreases with time, the first term $\propto
H^{4}(q-{1}/{2})$, with a lower power of $H$, will eventually
dominate and relax the CC in the matter era, where $q\simeq
{1}/{2}$. The arguments for the dynamical stabilization in $\rL$ are
exactly the same as in the toy model above. Another novelty is the
fixed parameter $y\sim H_{\text{eq}}^{-{2}/{3}}$, which is related
to the Hubble rate $H_{\text{eq}}\sim10^{5}\, H_{0}$ at the time of
matter-radiation equality, where~$H_0\sim 10^{-42}\,\text{GeV}$ is
the current value. In the late accelerating universe, $q<0$ and the
tiny expansion rate~$H\approx H_{0}$ makes $B\sim H^{4}$ the
relevant term. In summary, the smallness of the denominator $B$ in
$\rL$ during all cosmological epochs is the clue to the relaxation
mechanism. An additional physical argument in favor of it is the
growth of inhomogeneities, which is in accord with observations
\cite{Bauer2009}.

This is a good place to demonstrate the absence of fine-tuning in
our mechanism. Let us ignore~${\mathcal O}(1)$ factors in the
Friedmann equation $\rho_{c}=\rho_{m}+\CC+\beta/H^{4}$ at late
times. Since $\CC$ and the compensating term $\beta/H^{4}$ are much
larger than the matter density $\rho_{m}$ and the critical energy
density~$\rho_{c}$, we find, in very good approximation,
$H^{4}=|\beta/\CC|$ at late times. Obviously, small changes in the
parameter $\beta$ will induce only small changes in the Hubble
rate~$H$. Hence, it suffices to fix the order of magnitude of
$\beta$ to obtain a small $H\simeq H_0$. This has nothing to do with
fine tuning. For instance, take a typical GUT value
$|\rho_{\Lambda}|\simeq M_X^4$, with $M_X=10^{16}\,\text{GeV}$; then
$\beta=M^8$ with $M\sim 10^{-4}$ eV (of order of a light neutrino
mass) would do. An essential result thus follows: large $|\CC|$
guarantees small asymptotic $H\simeq H_0$ for reasonable $M$.

\section{\large\bf An action functional formulation of the relaxation mechanism}

Next we implement the CC relaxation mechanism, here for the first
time in the modified gravity setup and in the metric formalism
\cite{SotiriouFaraoni08,Carroll04}. The crucial part of the model is
a function $F(R,\GB)$ of the Ricci scalar $R$ and the Gau\ss-Bonnet
invariant~$\GB$. The complete action functional of our cosmological
model is the following:
\begin{equation} \mathcal{S}=\int
d^{4}x\,\sqrt{|g|}\left[\frac{R}{16\pi G_{N}}-\CC-\beta
F(R,\GB)+\mathcal{L}_{\text{mat}}\right],\label{eq:CC-Relax-action}
\end{equation}
where $\mathcal{L}_{\text{mat}}$ is the matter Lagrangian, and $\CC$
an arbitrarily large cosmological constant. $G_{N}$ is the Newton
constant and the parameter~$\beta$ will also be dimensionful in
general. The variational principle $\delta\mathcal{S}/\delta
g^{ab}=0$ then leads straightforwardly to the Einstein equations,
\begin{equation}
G_{ab}= -8\pi G_{N}\left[g_{ab}\,\CC+2\beta
E_{ab}+T_{ab}\right],\label{eq:Mod-Einstein-Eqs}
\end{equation}
involving the Einstein tensor~$G_{ab}=R_{ab}-\frac{1}{2}g_{ab}R$ and
the energy-momentum tensor~$T_{ab}$ of matter. Additionally, there
is a new tensor $E_{ab}$ coming solely from the $F(R,\GB)$ term in
the action~(\ref{eq:CC-Relax-action}). On a spatially flat FLRW
background with line element $ds^{2}=dt^{2}-a^{2}(t)d\vec{x}^{\,2}$
and scale factor~$a(t)$, the tensor components in
Eq.~(\ref{eq:Mod-Einstein-Eqs}) are given by
$G_{\,\,0}^{0}=-3H^{2}$, $G_{\,\, j}^{i}=-\delta_{\,\,
j}^{i}(2\dot{H}+3H^{2})$ and\begin{eqnarray}
E_{\,\,0}^{0} & = & \left[\frac{1}{2}F(R,\GB)-3(\dot{H}+H^{2})F^{R}+3H\dot{F}^{R}\right.\nonumber \\
 & - & 12H^{2}(\dot{H}+H^{2})F^{\GB}+12H^{3}\dot{F}^{\GB}\biggr],\label{eq:E00}\\
E_{\,\, j}^{i} & = & \delta_{\,\, j}^{i}\left[\frac{1}{2}F(R,\GB)-(\dot{H}+3H^{2})F^{R}+2H\dot{F}^{R}+\ddot{F}^{R}\right.\nonumber \\
 & + & (\dot{H}+H^{2})(8H\dot{F}^{\GB}-12H^{2}F^{\GB})+4H^{2}\ddot{F}^{\GB}\biggr],\label{eq:Eij}\end{eqnarray}
where $F^{R}$ and $F^{\GB}$ are partial derivatives of $F$ with
respect to~$R$ and $\GB$, respectively, and $\dot{H}=-H^{2}(q+1)$.

The matter sector is described by the standard energy-momentum
tensor of an ideal fluid,
with proper energy density~$\rho=T_{\,\,0}^{0}$ and pressure
$p=-T_{\,\, i}^{i}/3$ (summed repeated indices). Similarly, the
$F(R,\GB)$-functional induces the effective energy density and
pressure
\begin{equation} \rho_{F}=2\beta
E_{\,\,0}^{0}\,\,\,\,\text{and}\,\,\,\, p_{F}=-2\,\beta\,E_{\,\,
i}^{i}/3\,.\label{eq:rho-pres_F}
\end{equation}
Since~$G_{ab}$ and~$E_{ab}$ are covariantly conserved, $T_{ab}$ is
conserved, too. Therefore, the Bianchi identity on the FLRW
background, $\dot{\rho}+3H(\rho+p)=0$, is valid for both matter
($\rho_m,p_m$) and $F$-components ($\rho_{F},p_{F})$, and hence also
for ($\rL,\, p_{\Lambda {\rm eff}})$, where $\rL\equiv\CC+\rho_{F}$
is the total energy density of the effective vacuum sector and
$\pL\equiv-\CC+p_{F}$ denotes the corresponding pressure. The
generalized Friedmann-Lema\^\i tre equations read:
\begin{eqnarray}
3H^{2} & = & 8\pi G_{N}(\rho_{m}+\rho_{r}+\rL),\label{eq:Einstein-H2}\\
H^{2}\left(q-{1}/{2}\right) & = & 4\pi
G_{N}(p_{r}+\pL)\,.\label{eq:Einstein-q-1-2}
\end{eqnarray}
They contain the energy densities of pressureless matter
$\rho_{m}=\rho_{m}^{0}a^{-3}$ and
radiation~$\rho_{r}=\rho_{r}^{0}a^{-4}$ with
pressure~$p_{r}=\rho_{r}/3$, and the effective vacuum components.

Motivated by the toy-model discussion above, the function $F$ in our
setup is taken in the simple form $F(R,\GB)=1/B$, with $B$ given in
Eq.~(\ref{eq:B(R,G)}). The key to the relaxation mechanism in this
functional framework is the observation that every derivative of
$F$ in the expressions (\ref{eq:E00}) and (\ref{eq:Eij}) yields
another factor $B^{-1}$. Thus the induced quantities $\rho_{F}$ and
$p_{F}$ adopt, in general, the structure
\begin{equation}
\rho_{F}=\frac{N_{1}(H,q,\dot{q})}{B^{3}},\,\,\,\,
p_{F}=\frac{N_{2}(H,q,\dot{q},\ddot{q})}{B^{4}},\label{eq:rho-pres-F}\end{equation}
where $N_{1,2}$ are functions not proportional to~$B$.
Correspondingly, the effective vacuum energy density adopts the form
$\rL=\CC+N_{1}/B^{3}$ and, in complete analogy with the toy-model
discussion, it can be small because the large $\CC$ can be
dynamically compensated by letting $B$ become sufficiently small
(but non-zero), i.e.\ $B\rightarrow0$ -- hereafter referred to as
``relaxation condition''.

Note that, at the equilibrium point (e.g.\ $q\simeq 1$ in the
radiation era), both terms in $\rL$ are almost equal to each other,
apart from opposite signs. Therefore, $\rho_{F}$ behaves
approximately as another (large) cosmological constant. Obviously,
$\rL$ is not constant in general, because
$\rho_F=\rho_F(H,q,\dot{q})$ is time-dependent. In addition, there
is a corresponding compensation of the pressure terms $p_{F}$ and
$-\CC$ in $\pL$. Finally, since the functions (\ref{eq:E00}) and
(\ref{eq:Eij}) differ from each other, the effective EOS,
$\omega=\pL/\rL$, will be in general a non-trivial function of time
or redshift. This should have phenomenological implications.

\section{\large\bf Analytical and numerical analysis of the model}

\begin{figure}
\noindent
\centering{\includegraphics[width=1\columnwidth]{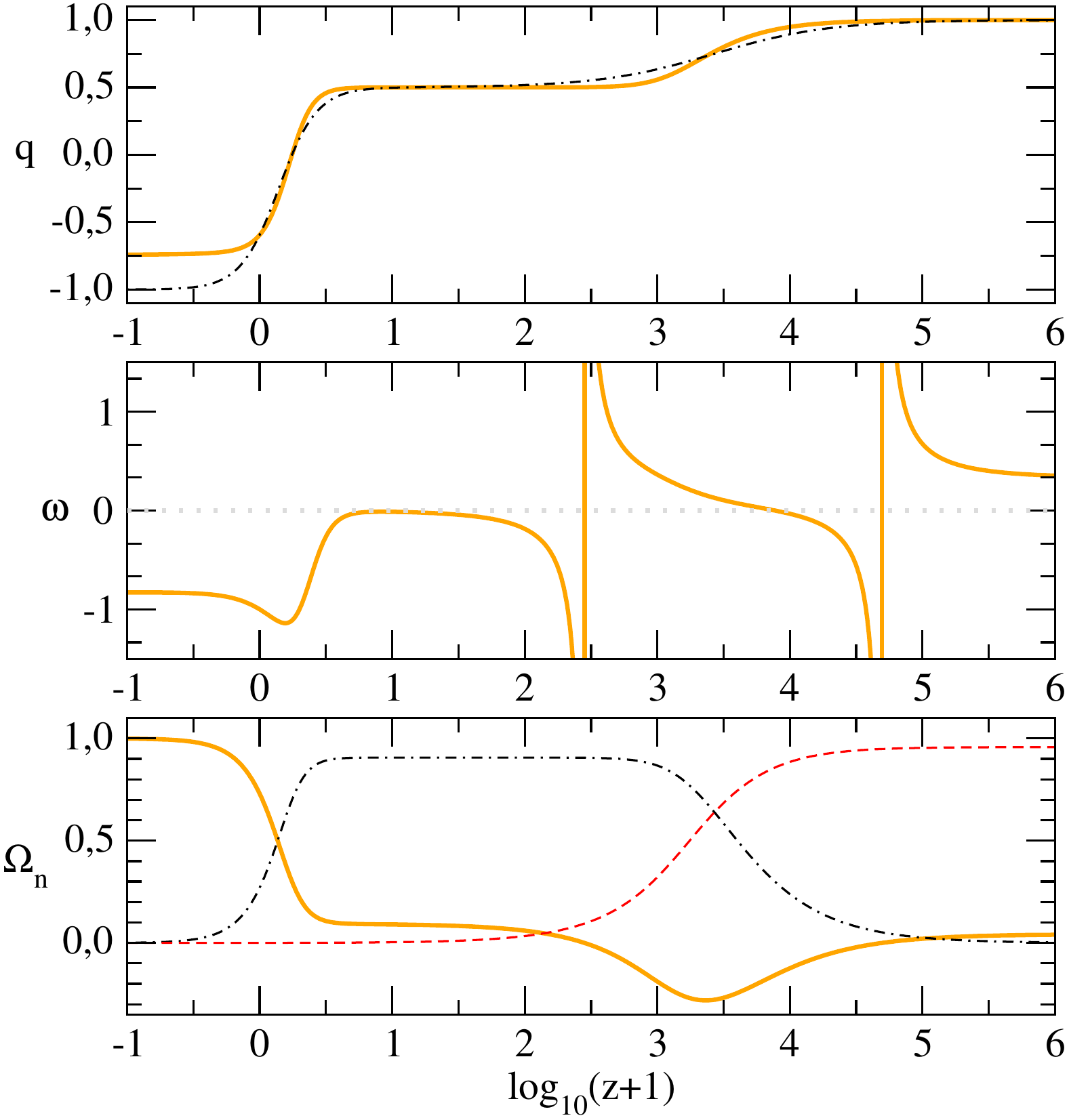}}
\caption{Deceleration parameter $q$, effective EOS~$\omega=\pL/\rL$,
and relative energy densities~$\Omega_{n}=\rho_{n}/\rho_{c}$ of the
effective vacuum~$\rL$ (orange thick curve), matter~$\rho_{m}$
(black dashed-dotted) and radiation~$\rho_{r}$ (red dashed) as
functions of the redshift~$z$. The thick orange curve in the $q$
plot corresponds to the relaxation model, and the black
dashed-dotted curve to $\Lambda$CDM. Model $F=1/B$, with inputs
$y=7\times 10^{-4}\, H_{0}^{-2/3}$, $\CC=-10^{60}\,\rho_{c}^{0}$,
$\Omega_{m}^{0}=0.27$, $\Omega_{r}^{0}=10^{-4}$, $q_{0}\approx-0.6$,
$\dot{q}_{0}=-0.5\,H_0$, with
$\rho_{c}^{0}=3H_0^2/(8\pi\,G_N)$.\label{fig:Model-1overB}}
\end{figure}

In the following, we will describe approximately the cosmic
evolution by making use of the CC relaxation condition, which is
equivalent to $|\rL|\ll|\CC|$. Remember that $B$ never
vanishes. Our analysis will be supported by exact numerical results
(cf.\ Fig.~\ref{fig:Model-1overB}) obtained by solving directly the
Friedmann equation (\ref{eq:Einstein-H2}) with $\rho_{F}$ given in
Eq.~(\ref{eq:rho-pres_F}).

Let us start in the matter era where $q\approx{1}/{2}$ and
$H^{2}\sim\rho_{m}\propto a^{-3}$. Applying the relaxation condition
in Eq.~(\ref{eq:B(H,q)}) leads to $(q-{1}/{2})\propto H^{2}$, and
Eq.~(\ref{eq:Einstein-q-1-2}) renders
\begin{equation}
H^{2}\left(q-{1}/{2}\right)=4\pi G_{N}(p_{r}+\pL)\propto
H^{4}\propto a^{-6}\,,\label{eq:MD-Friedmann}
\end{equation}
whereupon $\pL=-\rho_{r}/3+c_{1}\,a^{-6}$, with $c_{1}$ a constant;
and then via the Bianchi identity for ($\rL,\, p_{\Lambda {\rm
eff}})$ we find $\rL=c_{2}\,a^{-3}-\rho_{r}+c_{1}\,a^{-6}$, the
effective vacuum energy density, with $c_{2}$ another constant.
Thence
\begin{equation}
\omega=\frac{\pL}{\rL}=\frac{-\rho_{r}/3+c_{1}\,a^{-6}}{c_{2}\,a^{-3}-\rho_{r}+c_{1}\,a^{-6}}
\label{eq:EOS-MD}
\end{equation}
is the corresponding EOS, which interpolates between dust matter
($\omega\rightarrow0$) at late times and radiation
($\omega\rightarrow{1}/{3}$) in the early matter era. Depending on
the integration constant $c_{2}$ a pole might occur in $\omega$ when
$\rL$ changes its sign, see Fig.~\ref{fig:Model-1overB}. It is worth
noticing that $\rL$ behaves like dark matter in this epoch, and we
could speculate on incorporating both dark matter and dark energy
into $\rL$. Apart from that, we remark the approximate tracking
relation $\rL\propto\rho_{m}$,
which can be considered a cornerstone for solving the coincidence problem.%

We expect that the relaxation mechanism became active only as of the
early radiation era, because $F=1/B\sim H^{-6}$ in
Eq.~(\ref{eq:CC-Relax-action}) was sub-dominant during the preceding
periods of inflation and reheating, where $q$ was far from $1$.
However, deep in the radiation epoch, before the matter era, we find
$q\approx1$ and $F$ becomes very important. Then
$H^{2}\sim\rho_{r}\propto a^{-4}$ implies $R=6H^{2}(1-q)\propto
H^{\frac{4}{3}}\propto a^{-\frac{8}{3}}$ upon using the relaxation
condition in Eq.~(\ref{eq:B(H,q)}). Moreover,
Eqs.~(\ref{eq:Einstein-H2}) and (\ref{eq:Einstein-q-1-2}) lead to
the relation
\begin{equation} {3H^{2}(q-1)}={4\pi
G_{N}}\left[3\,\pL-\left(\rL+\rho_{m}\right)\right]\propto
a^{-{8}/{3}}\,,\label{eq:RD-Friedmann}\end{equation}
and thus to $\pL=(\rho_{m}+\rL)/3+c_{3}\,a^{-{8}/{3}}$, for constant
$c_{3}$. From here the Bianchi identity determines
$\rL=c_{4}\,a^{-4}-\rho_{m}-({9}/{4})\,c_{3}\,a^{-{8}/{3}}$, with a
new a constant $c_{4}$. Accordingly, the EOS interpolates between
radiation at early times and dust matter at later times:
\begin{equation} \omega=\frac{\pL}{\rL}=\frac{1}{3}\
\frac{1+({3c_{3}}/{4c_{4}})\,
a^{{4}/{3}}}{1-({\rho_{m}^{0}}/{c_{4}})\, a-({9c_{3}}/{4c_{4}})\,
a^{{4}/{3}}}.\label{eq:EOS-RD}
\end{equation}
Again a pole in $\omega$ (visible in Fig.~\ref{fig:Model-1overB})
can appear depending on $c_{4}$. As in the case of the matter epoch,
it entails no physical singularity since all the energy densities
remain finite; it merely reflects the change of sign of $\rL$.
Remarkably enough, we encounter the tracking
property~$\rL\propto\rho_{r}\propto a^{-4}$ also in the radiation
era.

At late times, the universe leaves the matter epoch ($\rho_m\sim
1/a^{3}\to 0$) and gradually enters the ``dark energy (DE) era''. As $q$
departs now significantly from~${1}/{2}$ or $1$, the last dynamical
resource left to the relaxation mechanism for compensating the big
$\rho_{\Lambda}$, is to choose a very low value of $H$, whereby the
DE epoch takes over somewhere near our time (cf.\
Fig.~\ref{fig:Model-1overB}) -- ``cosmic coincidence''. As the
lowest power of $H$ becomes dominant in Eq.~(\ref{eq:B(H,q)}), we
have $B\simeq 24H^{4}(q-{1}/{2})(q-2)$ and the full expression for
$\rho_{F}$ in this regime can be computed in closed form:
\begin{equation}
\frac{\rho_{F}}{\beta}=\frac{5q^{2}-3q-5}{2H^{4}(2q^{2}-5q+2)^{2}}-\frac{\dot{q}\,(4q^{2}-10q+7)}{2H^{5}(2q^{2}-5q+2)^{3}}.\label{eq:Late-rho-F}
\end{equation}
This equation holds for the recent, current and asymptotic future
times, i.e.\ whenever the universe is DE or matter dominated. With
$H\rightarrow0$, the previous expression would grow without limit,
unless the numerators eventually tend to zero, too. Thus, the
accelerating asymptotic solution ensues for $\dot{q}\rightarrow0$
and $q\rightarrow (3-\sqrt{109})/10\approx-0.74$, see
Fig.~\ref{fig:Model-1overB}. There are actually three solutions for
$t\rightarrow\infty$: a phantom phase, a de Sitter phase with
constant $H$ and an asymptotic power-law expansion $a(t)\propto
t^{r}$ with $r>1$. The previously found solution is of this sort,
with $r\rightarrow1/(q+1)\approx3.91$.

It is not difficult to work out generalized relaxation models, e.g.\
of the form $F=R^{n}/B^{m}$ with $m,n>0$. The crucial ingredient is
the function $B$ in the denominator of $F$. Generally, this yields
induced terms $\rho_{F},p_{F}\propto B^{-s}$ with $s>0$ as a result
of Eqs.~(\ref{eq:E00},\ref{eq:Eij}), where derivatives of $F$
introduce more factors of $B$ in the denominators. One can readily
show that the CC relaxation and tracking properties in the matter
and radiation eras follow again from the relaxation condition.
Furthermore,
the parameter~$\beta$ is related to a power of a mass scale $M$
as~$|\beta|=M^{4-2n+4m}$. The size of $M$ can be estimated by
applying the approximations $H,\dot{q}\sim H_{0}$, $q\in[-1,0]$,
together with the condition $|\CC|\simeq |\rho_{F}|\simeq
|\beta\,F|$, at $H\sim H_0$. Again we take $|\CC|\sim M_X^{4}$ for
the standard GUT energy density ($M_X=10^{16}\,\text{GeV}$). In the case
$(n,m)=(0,1)$, we recover the previous result
$M^{8}=|\beta|={\mathcal O}(\CC H_{0}^{4})$, i.e.\
$M\sim10^{-4}\,\text{eV}$. Similarly, we obtain $M\sim100$
MeV$\,\sim \Lambda_{\rm QCD}$, for $(n,m)=(3,2)$; and $M\sim M_X$,
for $(n,m)=(2,1)$. Remarkably, the mass parameter~$M$ of the
relaxation mechanism is not only completely free from fine-tuning
problems, it also lies in a perfectly reasonable range of particle
physics masses, possibly related to neutrinos, QCD or even GUT
models.

\section{\large\bf Conclusions}

To summarize, in this letter, we have unveiled a whole class of
modified gravity action models which, despite holding an arbitrarily
large CC at \textit{all} times since the early epochs, display a
very small ``effective CC'' at present, and without ever needing
fine tuning.
Like many modified gravity models our approach is not complete because of the existence of extra degrees of freedom and related problems, which require further considerations, see e.g.~\cite{Carroll04,Hindawi:1995cu}. Nevertheless, the {\em Relaxed Universe} scenario is {\em not} just another example for inducing late time cosmic acceleration, for the gravity modifications needed to dynamically compensate the large CC are crucial during the entire cosmic history.
The mechanism also predicts remarkable tracking properties,
and characteristic dynamical features in our recent past, which
alleviate the coincidence problem. A more detailed account of this
framework will be presented elsewhere.

\vspace{1cm}

\textbf{Acknowledgments}\hspace*{1cm} The authors have been supported by DIUE/CUR Generalitat de Catalunya under project 2009SGR502; FB and JS also by MEC and FEDER under project FPA2007-66665 and by the
Consolider-Ingenio 2010 program CPAN CSD2007-00042, and HS also by
the Ministry of Education, Science and Sports of the Republic of
Croatia under contract No. 098-0982930-2864.


\newpage

\begin{thebibliography}{50}

\bibitem{weinberg89} S.~Weinberg, Rev. Mod. Phys. {\bf 61} {(1989)} {1}.

\bibitem{CCproblem} P.J.E.~Peebles and B.~Ratra, Rev. Mod. Phys. {\bf 75} {(2003)} {559};
T.~Padmanabhan, Phys. Rept. {\bf 380} {(2003)} {235}; \, V. Sahni, A.
Starobinsky, Int. J. Mod. Phys. {\bf A9} {(2000)} {373}; S.M. Carroll,
Living Rev. Rel. {\bf 4} (2001) 1; E.J. Copeland, M. Sami, S.
Tsujikawa, Int. J. Mod. Phys. {\bf D15} {(2006)} {1753}, and
references therein.

\bibitem{Demir} D. A. Demir, Found. Phys. {\bf 39} (2009) 1407.

\bibitem{ShapSol09} I. L. Shapiro,  J. Sol\`a,  Phys. Lett. {\bf B682} (2009)
105.

\bibitem{LXCDM} J.~Grande, J.~Sol\`a, H.~\v{S}tefan\v{c}i\'{c},
JCAP 0608:011 (2006).

\bibitem{Stefancic08} H.~\v{S}tefan\v{c}i\'{c}, Phys. Lett. {\bf B670} {(2009)} {246}.

\bibitem{Odin} S.~Nojiri, S.~D.~Odintsov, Phys.\ Rev.\ D {\bf 72} (2005) 023003.

\bibitem{BSS09a} F.~Bauer, J.~Sol\`{a}, H.~\v{S}tefan\v{c}i\'{c}, Phys. Lett.  {\bf B 678} (2009)
427, arXiv:0902.2215.

\bibitem{Bauer2009} F.~Bauer, Class.\ Quant.\ Grav.\  {\bf 27} (2010) 055001, arXiv:0909.2237.

\bibitem{SotiriouFaraoni08}
  S. Nojiri and S.D. Odintsov,
  Int.\ J.\ Geom.\ Meth.\ Mod.\ Phys.\  {\bf 4} (2007) 115;
 T.P. Sotiriou, V. Faraoni, arXiv:0805.1726 [gr-qc];
R. Woodard, \textit{Lect. Notes  Phys.} 720 (2007) 403.

\bibitem{Carroll04} S.M. Carroll, V. Duvvuri, M. Trodden, M.S. Turner,  Phys. Rev. {\bf D70} {(2004)} {043528};
S. Nojiri, S.D. Odintsov, Phys. Rev. {\bf D68} {(2003)} {123512};
S.M. Carroll, A. de Felice, V. Duvvuri, D.A. Easson, M. Trodden, M.S.
Turner, Phys. Rev. {\bf D71} {(2005)} 063513.

\bibitem{Hindawi:1995cu} A.~Hindawi, B.A.~Ovrut, D.~Waldram, Phys.\ Rev.\ {\bf D53} (1996) 5597; A.~De Felice, M.~Hindmarsh, M.~Trodden, JCAP {\bf 0608} (2006) 005.

\end{thebibliography}
\end{document}